\documentclass[twocolumn,preprintnumbers,amsmath,amssymb,showpacs]{revtex4}
\usepackage{epsfig}
\begin{document}
\title{Tricritical behavior of the massive chiral Gross-Neveu model}
\author{Christian Boehmer}
\author{Michael Thies}
\author{Konrad Urlichs}
\affiliation{Institut f\"ur Theoretische Physik III,
Universit\"at Erlangen-N\"urnberg, D-91058 Erlangen, Germany}
\date{\today}
\begin{abstract}
The phase diagram of the massive chiral Gross-Neveu model (the 1+1-dimensional Nambu--Jona-Lasinio model at large $N$)
is investigated in the vicinity of the tricritical point. Using the derivative expansion, the grand canonical potential
is cast into the form of a Ginzburg-Landau
effective action. Minimization of this action by variational and numerical methods reveals both 1st and 
2nd order phase transitions to a chiral crystal phase, separated by a tricritical line. These findings are 
contrasted to the massive Gross-Neveu model with discrete chiral symmetry where only 2nd order transitions
have been observed.  
\end{abstract}
\maketitle

\section{Introduction}

Following Gross and Neveu \cite{1}, we consider four-fermion interaction models in 1+1 dimensions 
with Lagrangians
\begin{equation}
{\cal L}  = {\rm i} \bar{\psi}\partial \!\!\!/ \psi -m_0 \bar{\psi} \psi
+ \frac{g^2}{2} \left\{ \begin{array}{ll}  (\bar{\psi}\psi)^2 & {\rm (GN)} \\
 (\bar{\psi}\psi)^2 -(\bar{\psi}\gamma_5 \psi)^2 &{\rm (NJL)}\end{array} \right.
\label{a1}
\end{equation}
where we suppress flavor indices ($i=1...N$). Only the large $N$ limit is of interest to us,
so that the classic no-go theorems \cite{1a,1b} are bypassed.
For vanishing bare mass $m_0=0$,
the first variant (referred to as Gross-Neveu (GN) model hereafter) has a discrete chiral symmetry $\psi\to \gamma_5 \psi$,
the second one (the two-dimensional Nambu--Jona-Lasinio (NJL) model \cite{2}) a continuous chiral symmetry 
$\psi\to \exp({\rm i}\gamma_5 \alpha)\psi$.
The respective symmetry groups Z$_2$ and U(1) manifest themselves in the bound state 
spectrum. Thus the GN model possesses massive kink and kink-antikink baryons, where the
kink reflects most directly the underlying Z$_2$ symmetry \cite{3}. The NJL model features a pseudoscalar meson
and a baryon both of which become massless in the chiral limit due to chiral symmetry breaking \cite{4,5}. Whereas the meson (``pion") is
a small fluctuation in the flat direction of the vacuum familiar from the Goldstone theorem in higher dimensions,
the massless baryon is a more exotic object specific for 1+1 dimensions. It may be associated with a full turn around the
chiral circle. Since this happens ``adiabatically" as a function of $x$, it does not cost any extra energy
and the baryon mass vanishes.
One can show that  the winding number of the chiral field is baryon number \cite{4}. Hence the NJL baryon has a topological 
structure, reminiscent of the Skyrme model in 3+1 dimensions \cite{6}.

In the present work, we study the 1+1 dimensional NJL model at finite temperature $T$ and chemical potential $\mu$. The main issue
is the behavior of the chiral condensates
\begin{equation}
S-m_0 =-g^2\langle \bar{\psi}\psi\rangle, \qquad P=-g^2\langle \bar{\psi} {\rm i}\gamma_5\psi\rangle,
\label{a2}
\end{equation}
now regarded as thermal expectation values, as a function of $T$ and $\mu$. If one
enforces translational invariance and considers only homogeneous condensates as was done originally, the GN and
NJL models appear to have identical phase diagrams \cite{7,8}. From the physics point of view, this is puzzling, given
that the difference in chiral symmetries has such a dramatic impact on the structure and mass of individual
baryons.  It was realized during the last few years that GN models at finite chemical potential may develop an $x$-dependent
condensate, leading to solitonic crystal phases (for a recent review, see \cite{9}). If one takes into account these crystal phases, the phase
diagrams for the models with discrete and continuous chiral symmetry become quite different and the puzzle disappears.

So far, the phase diagrams of the massless and massive GN models have been worked out in full detail, to a large
extent analytically \cite{10,11}. By contrast, thermodynamics of the NJL model including the crystal phase has only been explored 
in the chiral limit \cite{12} where it leads to the phase diagram shown in Fig.~1.  Here, the horizontal line
$T=T_{\rm c}={\rm e}^{\rm C}/\pi$ (${\rm C}\approx 0.5772$ is the Euler constant)
represents a 2nd order critical line in the ($\mu,T$) diagram.  Above
this line, chiral symmetry is restored and the system behaves like an ideal massless Fermi gas. Below the line, chiral symmetry and
translational invariance are broken. The condensate exhibits a helical shape
(dubbed ``chiral spiral" in \cite{5}),
\begin{equation}
S- {\rm i}P = m {\rm e}^{2{\rm i}\mu x} ,
\label{a3}
\end{equation}
with a radius $m$ depending only on $T$ and a pitch equal to $\pi/\mu$. This phase
diagram has been derived in \cite{12} by ``gauging away" the chemical potential through a local $\gamma_5$-transformation. Since this
transforms the helical condensate (\ref{a3}) back onto the standard mass $m$, the $\mu$-independent phase boundary
shown in Fig.~1 follows at once. Unfortunately, this trick does not work at finite bare fermion mass $m_0$, and nothing is known
yet about the phase structure of the massive NJL$_2$ model. From Fig.~1 it is not obvious how the phase diagram will evolve
as we turn on the bare mass.
\begin{figure}
\begin{center}
\epsfig{file=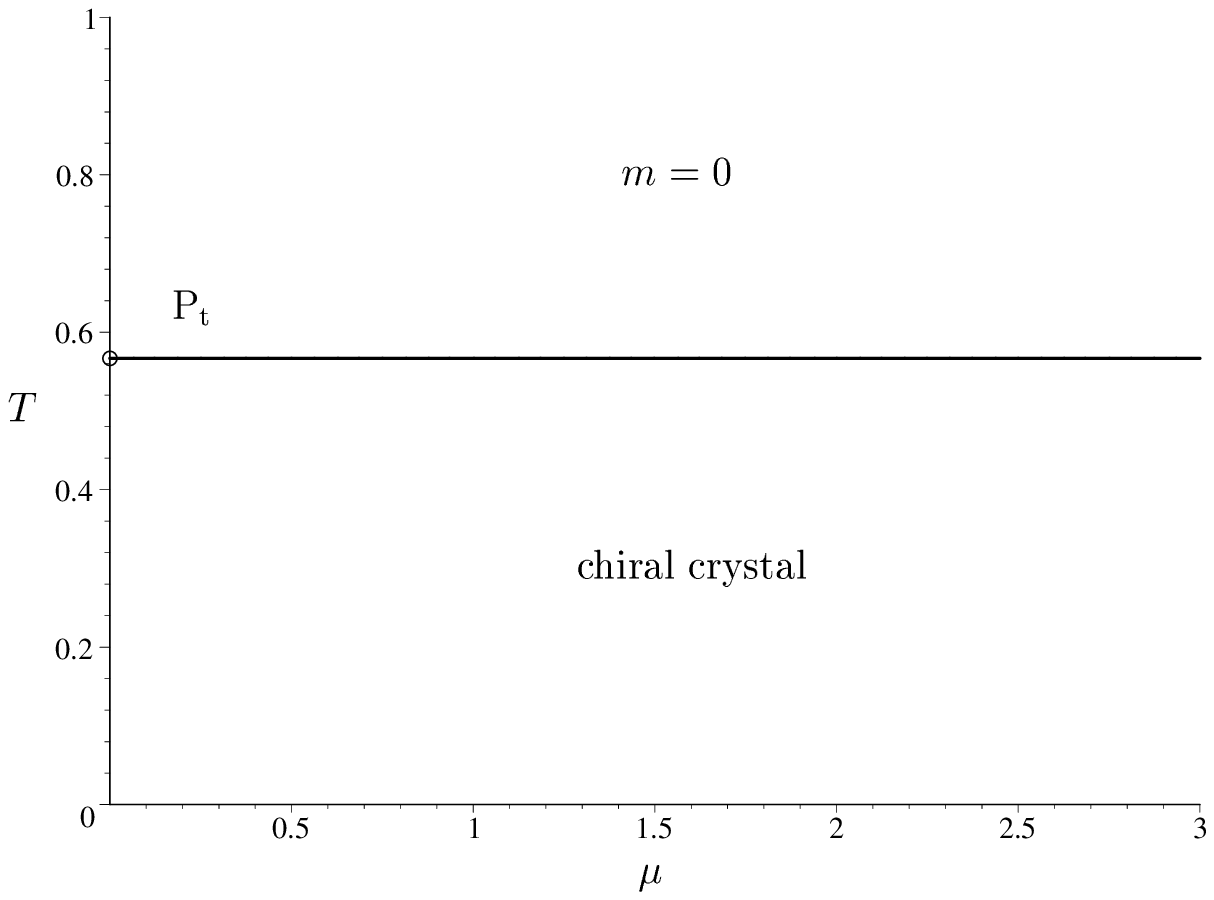,width=6.0cm,angle=270}
\end{center}
\caption{Phase diagram of the massless NJL$_2$ model in the ($\mu,T$) plane according to Ref.~\cite{12}. In the chiral crystal phase, the
condensate has helical symmetry, see Eq.~(\ref{a3}).}
\end{figure}

As a first step towards clarifying the phase structure of the massive NJL model, we propose to zoom in onto the tricritical
point and its neighborhood. The tricritical point is the point ${\rm P}_{\rm t}$ at ($\mu=0, T=T_{\rm c}$) where the horizontal
critical line crosses the vertical line $\mu=0$. The segment of this latter line between $T=0$ and $T=T_{\rm c}$ should also be
viewed as a critical line. It lies at the boundary of the phase diagram just because the baryons are massless (the critical
chemical potential at $T=0$ coincides with the baryon mass). The identification of ${\rm P}_{\rm t}$ as tricritical point will be fully
justified in the course of our calculations where we shall construct a tricritical line ending at this point.  

Finite $m_0$ introduces a new axis into the phase diagram which we label by
\begin{equation}
\gamma= \frac{\pi}{Ng^2} \frac{m_0}{m},
\label{a3a}
\end{equation}
following earlier work on the GN model \cite{9,11}. This is a dimensionless, renormalization group invariant
parameter which replaces the bare fermion
mass $m_0$ in physical observables. If we can understand
the vicinity of the point ${\rm P}_{\rm t}$ in ($\gamma,\mu,T$) space, we have a good chance to get basic
information about the phase diagram such
as orders of phase transitions and characteristics of different phases.
On the other hand, the calculations in this region are significantly easier because the condensates are weak and slowly varying.
This will enable us to avoid the tedious numerical Hartree-Fock (HF) procedure and solve instead 
a Ginzburg-Landau (GL) type theory where the fermions have been ``integrated out", leaving behind an effective
action for one complex boson field. Although such a calculation is approximate, it will become better and better
as we approach the tricritical point. In particular it will enable us to determine rigorously the asymptotics of how the phase 
boundaries approach the tricritical point. It therefore has the potential of providing relevant and well-defined information
about the phase structure of the NJL model, albeit in a limited region of the phase diagram. 

This paper is organized as follows. In Sect.~II, we outline the derivation of the GL effective action.
In Sect.~III, we minimize this action approximately by means of a simple variational calculation to gain some analytical
insight. In Sect.~IV we construct the phase diagram in the vicinity of ${\rm P}_{\rm t}$ by means of a full numerical solution
of the Euler-Lagrange equation. Sect.~V contains a comparison between the tricritical behavior of the GN and NJL models
and our conclusions.  
 
\section{Ginzburg-Landau theory near the tricritical point via the derivative expansion}

As in previous works we use the language of the relativistic HF approximation to implement the 
large $N$ limit \cite{12}. The single particle Dirac Hamiltonian with scalar and pseudoscalar mean fields $S,P$
reads
\begin{equation}
H=-\gamma^5 {\rm i} \partial_x + S(x) \gamma^0 + P(x){\rm i} \gamma^1,
\label{e1}
\end{equation}
where the $\gamma$-matrices can be chosen as
\begin{equation}
\gamma^0  =  \sigma_1 , \qquad
\gamma^1  =  -{\rm i}\sigma_2, \qquad 
\gamma^5  = \sigma_3.
\label{e2}
\end{equation}
In HF approximation the grand canonical potential consists of a single particle contribution plus a double counting correction,
\begin{eqnarray}
\Psi & = &  - \frac{1}{\beta} \int {\rm d}E \sigma(E) \ln \left( 1+{\rm e}^{-\beta(E-\mu)}  \right) 
\nonumber \\
& & + \int {\rm d}x  \frac{(S-m_0)^2+P^2}{2Ng^2},
\label{e3}
\end{eqnarray}
and has to be minimized with respect to $S$ and $P$. The complicated part is the dependence of  
the spectral density $\sigma(E)$ on $S$ and $P$. The spectral density
\begin{equation}
\sigma(E)= {\rm Tr} \delta(H-E) = \frac{1}{\pi} {\rm Im} R(E+{\rm i}\epsilon)
\label{e4}
\end{equation}
is closely related to the resolvent of $H$,
\begin{equation}
R(z) = {\rm Tr} \frac{1}{H-z}= {\rm Tr}\frac{H+z}{H^2-z^2}.
\label{e5}
\end{equation}
If the mean fields are weak and slowly varying, we can apply the derivative expansion \cite{13}
to derive an approximate, closed expression for $\sigma(E)$ and hence for $\Psi$. It will 
involve a polynomial in $S,P$ and their derivatives and may be viewed as a microscopic GL theory
for the NJL model near the tricritical point.
The calculation is straightforward and has been done along the lines explained in more detail in \cite{14,15}.
We decompose $H$ and $H^2$ as follows, 
\begin{equation}
H=K+I, \qquad H^2 = H_0^2+V.
\label{e6}
\end{equation} 
The resolvent can then be expanded formally as
\begin{equation}
R(z)  =  {\rm Tr}(K+I+z)\left(G_0\sum_{n=0}^{\infty}(-VG_0)^n\right)
\label{e7}
\end{equation}
with
\begin{equation}
G_0=\frac{1}{H_0^2-z^2}.
\label{e8}
\end{equation}
We identify the unperturbed part $K$ of $H$ with the free massless Dirac Hamiltonian and similarly $H_0^2$ with $K^2$
and combine the mean fields $S,P$ into one complex field
\begin{equation}
\Phi  =  S -{\rm i}P .
\label{e9}
\end{equation}
This specifies all operators entering the derivative expansion (\ref{e7}) as follows, 
\begin{eqnarray}
K&=&\left( \begin{array}{cc} -{\rm i}\partial_x & 0 \\ 0 & {\rm i}\partial_x \end{array} \right),
\qquad
I \ = \ \left( \begin{array}{cc} 0  & \Phi  \\ \Phi^* & 0 \end{array} \right)
\nonumber \\
H_0^2 & = & -\partial_x^2 , \qquad
V \  = \   \left( \begin{array}{cc} |\Phi|^2 & -{\rm i}\Phi' \\ {\rm i}(\Phi')^* & |\Phi|^2 \end{array} \right).
\label{e10}
\end{eqnarray}
As in a previous application of this formalism to baryons \cite{15} all infinities
can be controlled with the help of the vacuum gap equation
\begin{equation}
\frac{(1-m_0)}{Ng^2} = \frac{1}{\pi} \ln \Lambda
\label{e11}
\end{equation}
and the following relation for the chiral symmetry breaking parameter
\begin{equation}
\gamma=m_0\ln \Lambda,
\label{e12}
\end{equation}
where we use units where the vacuum fermion mass is $m=1$ for all $\gamma$. 
Counting the field $\Phi$ and each derivative $\partial_x$ formally as being of order $\epsilon$, we
evaluate all terms in the expansion (\ref{e7}) up to order $\epsilon^6$. 
With the help of the momentum space technique developed in 
\cite{15}, we find after a lengthy calculation the effective action (equivalent to the grand canonical potential density)
\begin{eqnarray}
\Psi_{\rm eff} & = & \alpha_0 + \alpha_1 \left( |\Phi|^2-2 {\rm Re}\, \Phi \right) +\alpha_2 |\Phi|^2
+ \alpha_3 {\rm Im}\, \Phi (\Phi')^*
\nonumber \\
& + &  \!\!   \alpha_4 \!\! \left( |\Phi|^4 + |\Phi'|^2 \right)\! +\! \alpha_5 {\rm Im}\! \left( \Phi'' - 3 |\Phi|^2  \Phi \right)\! (\Phi')^{*} 
\label{e13} \\
&+ & \alpha_6 \left( 2|\Phi|^6 + 8 |\Phi|^2|\Phi'|^2+ 2{\rm Re}\, (\Phi')^2 (\Phi^*)^2 + |\Phi''|^2 \right)
\nonumber 
\end{eqnarray}
The coefficients $\alpha_i$ are $(\gamma,\mu,T)$-dependent functions 
\begin{eqnarray}
\alpha_0 & = & - \frac{\pi}{6} T^2 - \frac{\mu^2}{2\pi}-\frac{1}{4\pi}
\nonumber \\
\alpha_1 & = & \frac{\gamma}{2\pi}
\nonumber \\
\alpha_2 & = &  \frac{1}{2\pi} \left[ \ln (4\pi T) + {\rm Re}\, \psi \left(z\right) \right]
\nonumber \\
\alpha_3 & = &  - \frac{1}{2^3 \pi^2 T} {\rm Im}\, \psi \left(1,  z \right) 
\nonumber \\
\alpha_4 & = &  - \frac{1}{2^6 \pi^3 T^2} {\rm Re}\,  \psi \left(2, z\right)
\nonumber  \\
\alpha_5 & = & - \frac{1}{2^8 3 \pi^4 T^3} {\rm Im}\, \psi \left(3, z \right)
\nonumber \\
\alpha_6 & =  &  \frac{1}{2^{12}3 \pi^5 T^4} {\rm Re}\,  \psi \left(4, z \right)
\label{e14}
\end{eqnarray}
expressed through standard di- and polygamma functions of 
\begin{equation}
z=\frac{1}{2}+ \frac{{\rm i}\mu}{2\pi T}.
\label{e14a}
\end{equation}
For real $\Phi$, Eqs.~(\ref{e13}-\ref{e14a}) reproduce the effective action for the GN model which was derived
in \cite{11} from the full solution. The novel terms ($\sim \alpha_3,\alpha_5$) can be tested in the chiral
limit where the exact $\Phi$ is known from the chiral spiral \cite{12}. The derivative expansion is particularly useful
in a case like the NJL model where the exact answer is not known.

In the GN model, it was imperative to go to 6th order since the terms of order $\epsilon^2$ and
$\epsilon^4$ both vanish at the tricritical point. In the NJL model, this is not the case since the 
tricritical point is now located elsewhere. Here it is sufficient to keep the terms
up to 4th order ($\alpha_0$...$\alpha_4$) for a leading order calculation.
The original HF problem
is then reduced to a complex $\Phi^4$ theory which can be treated classically in the large 
$N$ limit ($\Phi$ is the HF potential, a classical quantity).
The Euler-Lagrange equation then becomes the (inhomogeneous) complex non-linear 
Schr\"odinger equation
\begin{equation}
\alpha_4 \Phi''-{\rm i}\alpha_3 \Phi' - (\alpha_1+\alpha_2+2\alpha_4|\Phi|^2)\Phi + \alpha_1=0.
\label{e15}
\end{equation}
The solution of this differential equation yields directly the HF potential, without need to solve the Dirac-HF equation
self-consistently. This drastic simplification only occurs in regions
of the phase diagram where the mean field is both weak and smooth.
Unfortunately, it does not seem possible to integrate Eq.~(\ref{e15}) in closed analytical form. 
Eventually, we shall solve it numerically in order to minimize the effective action.
Before turning to the results, we describe a semi-quantitative variational calculation.
It has all the salient features of the full numerical solution while being analytically tractable.

\section{A simple variational calculation}

In the present section we minimize the (4th order) GL effective action
\begin{eqnarray}
\Psi_{\rm eff}  & = &   \alpha_0 + \alpha_1 \left( |\Phi|^2-2 {\rm Re}\, \Phi \right) +\alpha_2 |\Phi|^2
+ \alpha_3 {\rm Im}\, \Phi (\Phi')^*
\nonumber \\
& &  + \alpha_4 \left( |\Phi|^4 + |\Phi'|^2 \right)
\label{e15a} 
\end{eqnarray}
using the following variational ansatz
\begin{equation}
\Phi = M + A {\rm e}^{{\rm i}Qx},
\label{e16}
\end{equation}
with three real parameters $M,A,Q$. The ansatz is motivated by the fact that it contains the homogeneous solution
($A=0$) and the chiral spiral ($M=0$) as special cases and by the desire to perform the whole calculation analytically.
Inserting Eq.~(\ref{e16}) into Eq.~(\ref{e15a}) and averaging over one 
period $L=2\pi/Q$ yields the effective potential
\begin{eqnarray}
\Psi_{\rm eff} &=& \alpha_0+\alpha_1 (M^2+A^2-2M)+ \alpha_2 (M^2+A^2)
\nonumber \\
&- &\!\! \alpha_3 A^2Q + \alpha_4 \!\left[M^4+A^4+A^2(4M^2+Q^2)\right] 
\label{e17}
\end{eqnarray}
This 4th order polynomial has to be minimized with respect to $M,A,Q$.
We first vary with respect to $Q$, 
\begin{equation}
\partial_Q \Psi_{\rm eff} = A^2\left(2\alpha_4 Q- \alpha_3 \right) = 0.
\label{e18}
\end{equation}
This condition yields either $A=0$, i.e., the homogeneous phase
\begin{equation}
\Psi_{\rm hom} = \alpha_0 + \alpha_1(M^2-2M)+\alpha_2 M^2 + \alpha_4 M^4,
\label{e19}
\end{equation}
or it determines the wavenumber $Q$ to be
\begin{equation}
Q=\frac{\alpha_3}{2\alpha_4}.
\label{e20}
\end{equation}
In the latter case, we arrive at the effective potential governing the crystal phase,
\begin{eqnarray}
\Psi_{\rm cryst} &=& \alpha_0 + \alpha_1 (M^2+A^2-2M)+\alpha_2 (M^2+A^2)
\nonumber \\
& & + \alpha_4(M^4+A^4+4A^2M^2)-\frac{\alpha_3^2}{4\alpha_4} A^2,
\label{e21}
\end{eqnarray}
which we minimize with respect to $A^2$ next,
\begin{equation}
\partial_{A^2} \Psi_{\rm cryst} = \alpha_1+\alpha_2 +2\alpha_4(A^2+2M^2)-\frac{\alpha_3^2}{4\alpha_4}=0
\label{e22}
\end{equation}
or
\begin{equation}
A^2=-2M^2-\frac{\alpha_1+\alpha_2}{2\alpha_4}+ \frac{\alpha_3^2}{8\alpha_4^2} .
\label{e23}
\end{equation}
A real value of $A$ is only obtained if 
\begin{equation}
M^2 \le \frac{\alpha_3^2}{16 \alpha_4^2}-\frac{\alpha_1+\alpha_2}{4\alpha_4} = M_{\rm max}^2.
\label{e24}
\end{equation}
Keeping in mind this upper bound for $M^2$, we eliminate $A^2$ from Eq.~(\ref{e21}) and get an effective potential of the 
single variable $M$ for the crystal phase,
\begin{eqnarray}
\Psi_{\rm cryst} &=& \alpha_0  - \frac{(\alpha_1+\alpha_2)^2}{4\alpha_4} + \frac{\alpha_3^2(\alpha_1+\alpha_2)}{8\alpha_4^2}-
\frac{\alpha_3^4}{64 \alpha_4^3}
\nonumber \\
&- & \!\!2\alpha_1 M \!- \!\left(\alpha_1+\alpha_2-\frac{\alpha_3^2}{2\alpha_4}\right)\!\!M^2 - 3 \alpha_4 M^4
\label{e25}
\end{eqnarray}
It is now possible to combine the information about both phases contained in Eqs.~(\ref{e19},\ref{e25}) into a single effective potential of $M$ as
follows. For $M^2>M_{\rm max}^2$, $\Psi_{\rm eff}$ is unique and given by $\Psi_{\rm hom}$, Eq.~(\ref{e19}).
For $M^2<M_{\rm max}^2$, $ \Psi_{\rm eff}$ is double valued, equalling either $\Psi_{\rm hom}$ or $\Psi_{\rm cryst}$. Actually, in this
region
\begin{equation}
\Psi_{\rm cryst}- \Psi_{\rm hom}  = -4\alpha_4(M^2-M_{\rm max}^2)^2\le 0.
\label{e26}
\end{equation}
Hence the homogeneous branch can simply be disregarded in the region $M^2\le M_{\rm max}^2$ when minimizing 
$\Psi_{\rm eff}$. The following effective potential thus contains all the pertinent information about the phase structure, 
\begin{equation}
\Psi_{\rm eff} = \Theta(M^2-M_{\rm max}^2) \Psi_{\hom} + \Theta(M_{\rm max}^2-M^2) \Psi_{\rm cryst}.
\label{e27}
\end{equation}
Although the definition is piecewise, the crystal and homogeneous effective potentials 
and their first derivatives with respect to $M$ match at $M=\pm M_{\rm max}$, so that our final effective potential
is a smooth function of $M$ (the 2nd derivative is discontinuous at $\pm M_{\rm max}$). This enables us to analyse the phase 
structure in a simple way. The system is in the crystal or the homogeneous phase depending of whether the global
minimum of $\Psi_{\rm eff}$ is in the region $M^2>M_{\rm max}^2$ or $M^2<M_{\rm max}^2$. A first order phase transition can be
identified by an effective potential  having two minima of equal depth, one on each side of $M_{\rm max}$. If a 2nd order
phase transition happens,
there is a unique minimum which crosses the boundary $M=M_{\rm max}$. Since $\Psi_{\rm eff}$ is
a 4th order polynomial in $M$, the minima can be found analytically by solving a cubic equation.

We now turn to the results. In Fig.~2, we show the phase boundary obtained at $\gamma=0.0005$,
restricting the ($\mu,T$) region to a neighborhood of the tricritical point.
This plot is representative for all $\gamma\ll1$.
\begin{figure}
\begin{center}
\epsfig{file=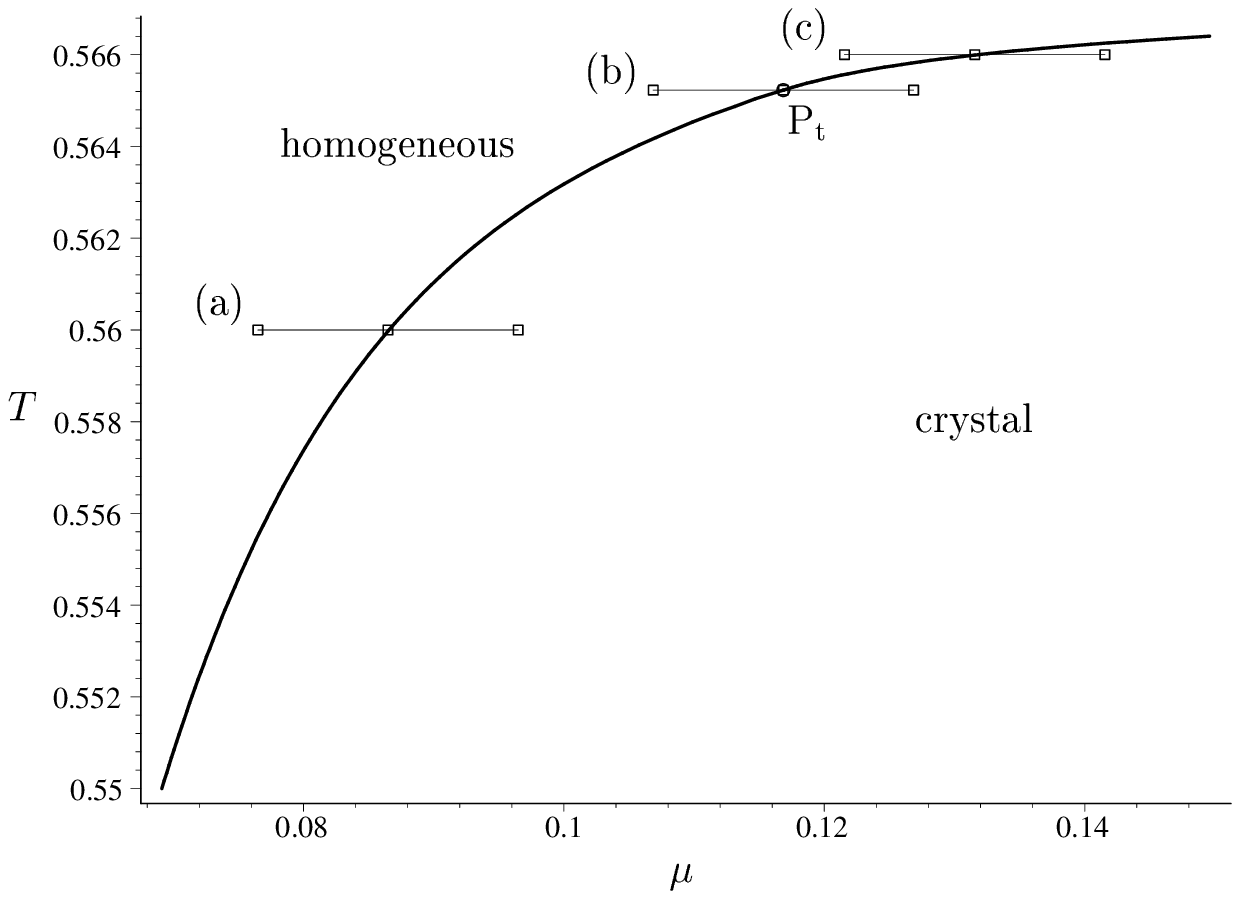,width=6.0cm,angle=270}
\end{center}
\caption{Phase boundary at $\gamma=0.0005$ according to the variational calculation based on the ansatz (\ref{e16}),
separating the homogeneous phase from the crystal phase.
A 1st order line to the left of the tricritical point ${\rm P}_{\rm t}$ goes over into a 2nd order line to the right of
${\rm P}_{\rm t}$. The points labelled (a,b,c) refer to the effective potentials shown in Fig.~3.}
\end{figure}
The most striking result is the appearance of a tricritical point ${\rm P}_{\rm t}$ separating a first order
transition line (to the left) from a 2nd order line (to the right). To illustrate the way in which the order was identified,
we show in Fig.~3 the effective potentials belonging to the points in Fig.~2 labeled by (a,b,c) and crossing the phase
boundary at three different temperatures.
\begin{figure}
\begin{center}
\epsfig{file=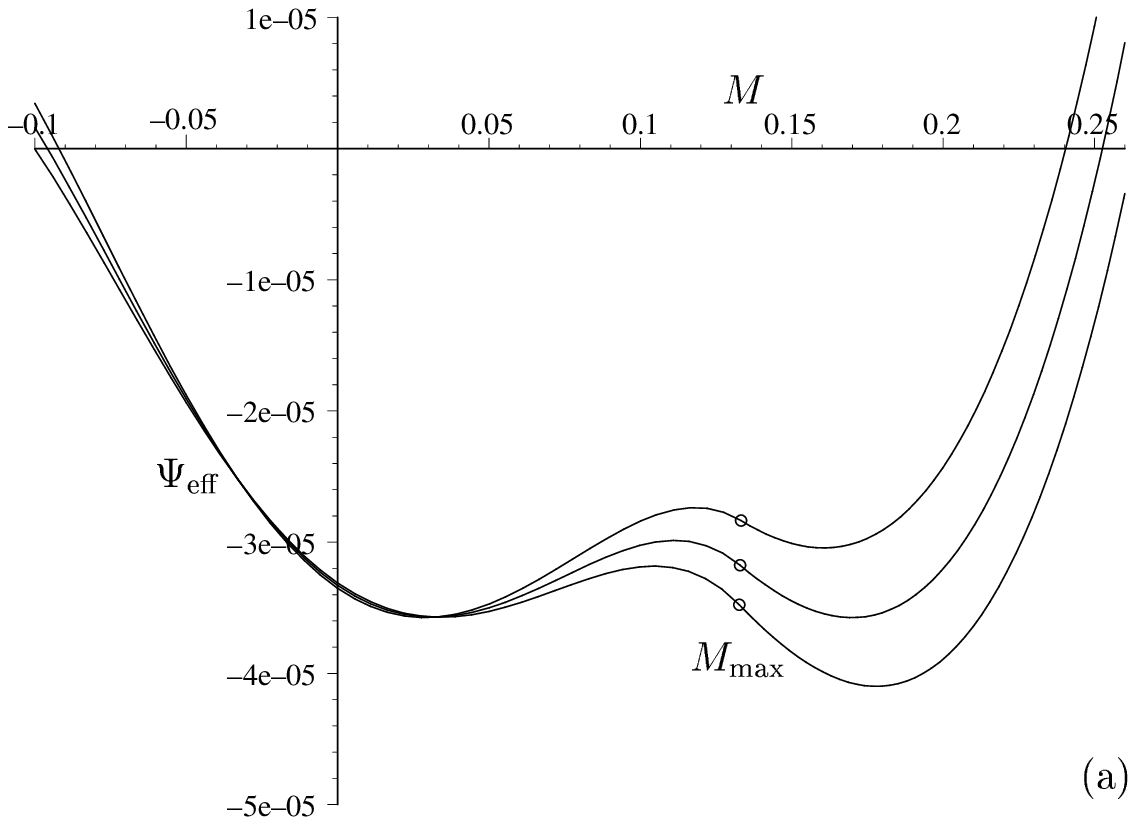,width=5.0cm,angle=270}\\ \vskip 0.2cm
\epsfig{file=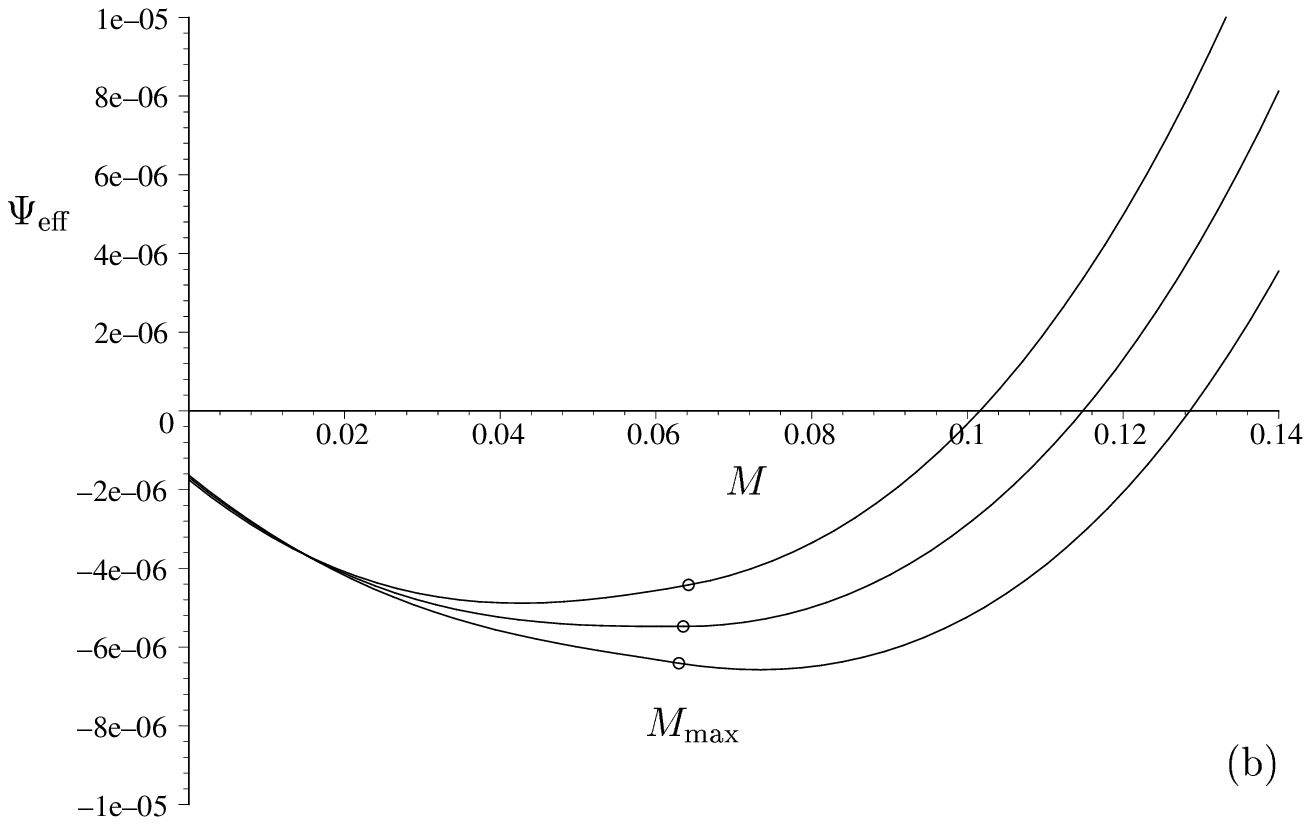,width=5.0cm,angle=270}\\ \vskip 0.2cm
\epsfig{file=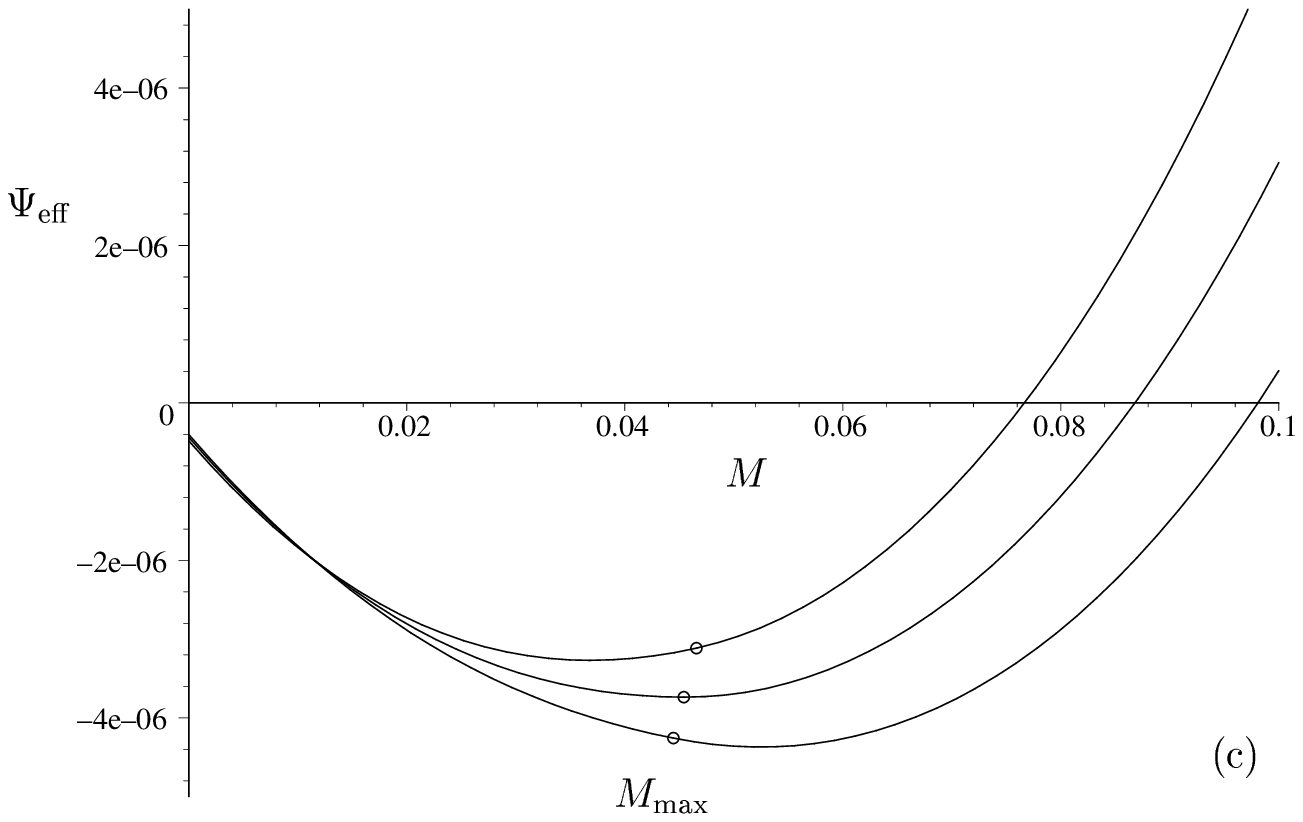,width=5.0cm,angle=270}
\end{center}
\caption{Evolution of effective potential when crossing the phase boundary in Fig.~2 at three different temperatures,
illustrating the difference between 1st and 2nd order transitions.
A minimum to the left of $M_{\rm max}$ refers to the crystal phase, a minimum to the right of 
$M_{\rm max}$ to the homogeneous phase. Figures 3a, 3b, 3c correspond to the points a,b,c marked in Fig.~2}
\end{figure}
As explained above, the effective potential contains crystal and homogeneous solution to the left and right of
$M_{\rm max}$. One clearly sees the difference between the first order transition
(a) and the 2nd order transition (c).
At the tricritical point (b), the 2nd derivative $\Psi_{\rm eff}''(M)$ to the left (i.e., in the crystal phase) vanishes
in addition to 1st derivative (remember that $\Psi_{\rm eff}''$ is discontinuous). Inspection of the effective potentials 
reveals that the instability can be described as follows: Above $T_{\rm t}$ when moving from
left to right through the phase boundary, there is a perturbative instability towards modulating the mean field,
\begin{equation}
\Phi=M \to \Phi=M+A{\rm e}^{{\rm i}Qx}
\label{e28}
\end{equation}
$M$ is continuous and $A$ vanishes at the phase boundary. This is the standard perturbative scenario also observed in the GN
model in the upper sheet of the phase diagram \cite{11}. Below $T_{\rm t}$, the mean field changes discontinuously
\begin{equation}
\Phi=M \to \Phi=M' + A{\rm}e^{{\rm i}Qx}
\label{e29}
\end{equation}
where $M\neq M'$ and $A$ jumps from 0 to a finite value. This is remarkably different from the GN model which shows
another type of instability here, namely towards formation of a single baryon in a 2nd order transition \cite{11}.
Repeating this calculation at several values of $\gamma$ finally yields the phase diagram shown in Fig.~4.
It has two sheets of 1st and 2nd order transitions separated by a tricritical line and nicely matches the phase diagram 
of Fig.~1 in the limit $\gamma \to 0$.
\begin{figure}
\begin{center}
\epsfig{file=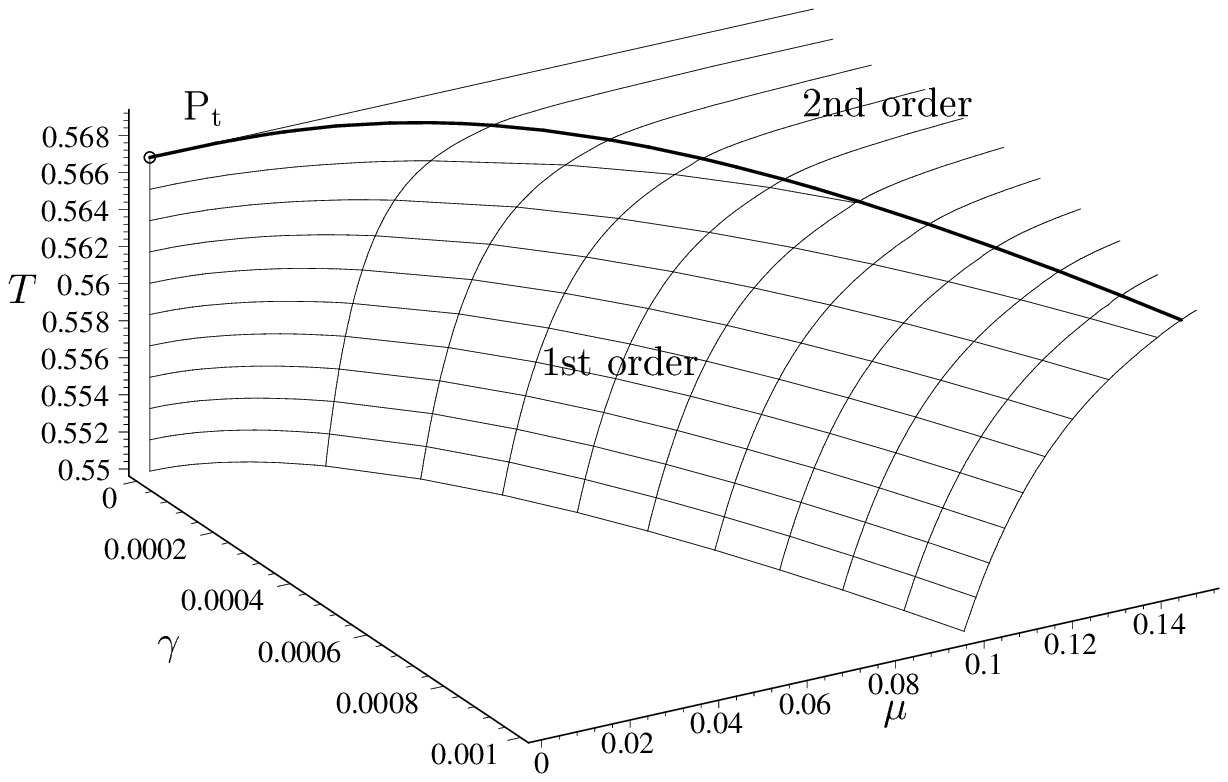,width=5.0cm,angle=270}
\end{center}
\caption{Two-dimensional phase boundary in ($\gamma,\mu,T$) space, separating a massive Fermi gas from the chiral crystal,
according to the variational calculation. Notice that only the vicinity of the tricritical point is shown. 
}
\end{figure}

Let us now try to understand how exactly the tricritical line approaches the tricritical point ${\rm P}_t=(0,0,T_c)$.
On general grounds one expects some power law behavior in variables which vanish at the phase transition. We 
therefore first replace the temperature by the shifted variable
\begin{equation}
\tau=\sqrt{T_c-T}.
\label{e30}
\end{equation}
In Figs.~5a and 5b, we show the phase boundaries for a sequence of $\gamma$-values in the ($\mu,T$)- plane
(a projection of the surface in Fig.~4 onto the ($\mu,T$)-plane) and in the ($\mu,\tau$)-plane.
\begin{figure}
\begin{center}
\epsfig{file=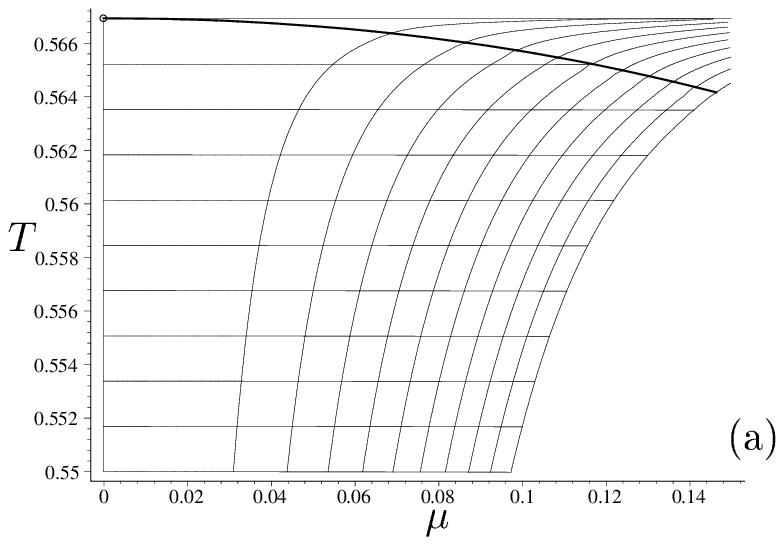,width=5.0cm,angle=270}\\ \vskip 0.2cm
\epsfig{file=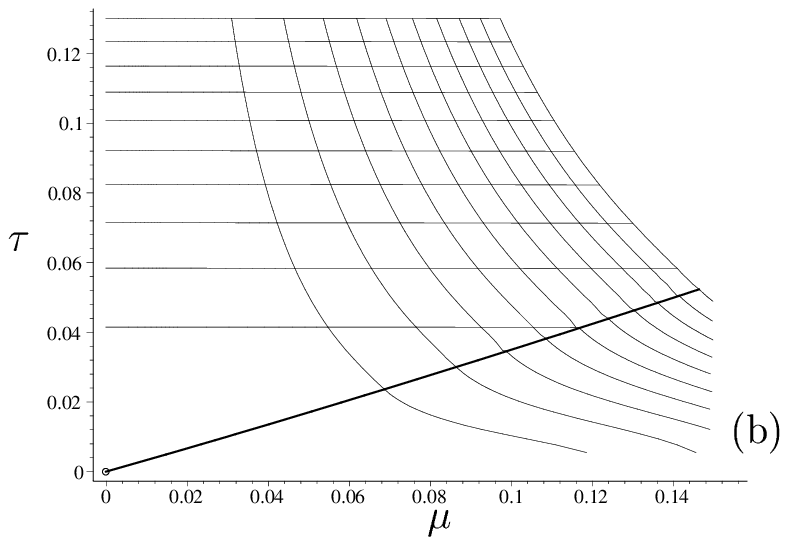,width=5.0cm,angle=270}\\ \vskip 0.2cm
\epsfig{file=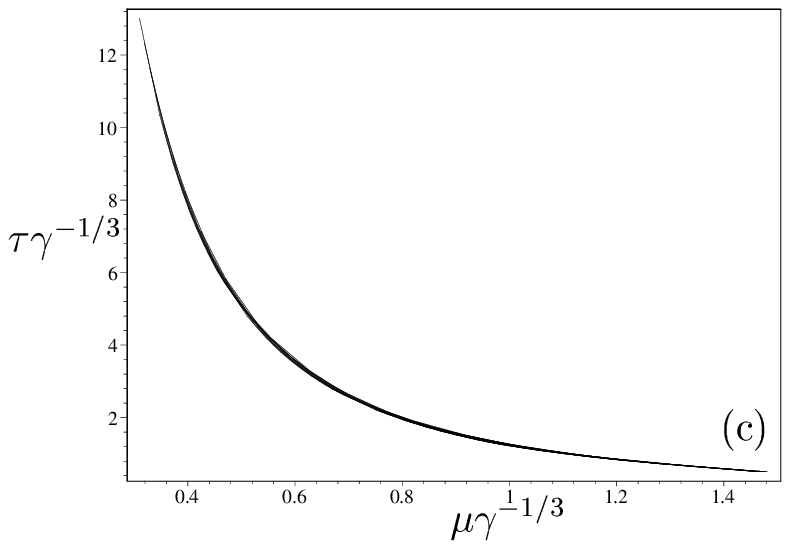,width=5.0cm,angle=270}
\end{center}
\caption{a) Projection of the 3d phase boundary of Fig.~4 onto the ($\mu,T$) plane. b) Same plot, replacing
the variable $T$ by $\tau=\sqrt{T_{\rm c}-T}$. c) Rescaling both axes by a common factor $\gamma^{-1/3}$
removes the $\gamma$-dependence.}
\end{figure}
This is not yet very illuminating. However, if we now rescale both axes in Fig.~5b
by a factor $\gamma^{-1/3}$, we obtain the striking 
result shown in Fig.~5c. We see that this particular rescaling removes the entire $\gamma$ dependence, at least within the
resolution of our plot. 
Thus the information contained in the 3d plot Fig.~4 could have been inferred from a computation at a single 
value of $\gamma$, e.g. the curve shown in Fig.~2. The tricritical line is just one point in the scaling diagram
located at $(\mu \gamma^{-1/3},\tau \gamma^{-1/3})\approx (1.488,0.488)$. In the next section, we show
that this scaling behavior has nothing to do with the specific variational ansatz, but can be derived
more generally from the effective action and simplifies the full calculation considerably.

\section{Tricritical behavior in the full calculation}

We first explain the reason behind the simple scaling law observed in the above variational calculation. 
We restrict ourselves to values of the independent variables ($\gamma,\mu,\tau$) of the order
\begin{equation}
\gamma  \sim   \epsilon^3, \qquad \mu \sim \epsilon,\qquad \tau =\sqrt{T_{\rm c}-T}\sim \epsilon.
\label{f1}
\end{equation}
This will be justified afterwards by showing that the tricritical line indeed lies in this region. We then
expand the coefficients $\alpha_i$, Eq.~(\ref{e14}), around the tricritical point,
\begin{eqnarray}
\alpha_1 & = & \frac{\gamma}{2\pi}
\nonumber \\
\alpha_2 & = & \frac{7}{8\pi}\zeta(3){\rm e}^{-2{\rm C}}\mu^2- \frac{1}{2} {\rm e}^{-{\rm C}}\tau^2 + {\rm O}(\epsilon^4)
\nonumber \\
\alpha_3 & = & \frac{7}{8\pi}\zeta(3){\rm e}^{-2{\rm C}}\mu + {\rm O}(\epsilon^3)
\nonumber \\
\alpha_4 & = & \frac{7}{32\pi}\zeta(3){\rm e}^{-2{\rm C}}+ {\rm O}(\epsilon^2)
\label{f2}
\end{eqnarray}
Returning to the Euler-Lagrange equation, Eq.~(\ref{e15}), dividing by $\alpha_4$ and using the expansion (\ref{f2}),
we find to leading order
\begin{equation}
\Phi''-4{\rm i}\mu \Phi' -\left(4 \mu^2- \frac{a}{T_{\rm c}}\tau^2 + 2 |\Phi|^2\right)\Phi + a  \gamma = 0
\label{f3}
\end{equation}
where we have introduced the transcendental constant
\begin{equation}
a = \frac{16 {\rm e}^{2 \rm C} }{7\zeta(3)} \approx 6.032
\label{f4}
\end{equation}
We next define a rescaled scalar field 
\begin{equation}
\Phi(x) = \gamma^{1/3} \varphi(\gamma^{1/3}x)
\label{f7}
\end{equation}
with
\begin{equation}
\Phi'(x) = \gamma^{2/3} \dot{\varphi}(\gamma^{1/3}x), \qquad \Phi''(x) = \gamma \ddot{\varphi}(\gamma^{1/3}x),
\label{f8}
\end{equation}
where $\dot{\varphi}$ denotes the derivative of $\varphi$ with respect to the argument $\xi= \gamma^{1/3}x$. 
Inserting this ansatz into the differential equation (\ref{f3}) and employing appropriately rescaled variables 
for chemical potential and temperature (the constant factors of O(1) are introduced to simplify the notation below),
\begin{equation}
\nu = 2 \gamma^{-1/3}\mu, \qquad \sigma = \sqrt{\frac{a}{T_{\rm c}}} \gamma^{-1/3} \tau,
\label{f9}
\end{equation}
the $\gamma$-dependence drops out,
\begin{equation}
\ddot{\varphi} - 2{\rm i}\nu \dot{\varphi} - \left(\nu^2-\sigma^2+2|\varphi|^2\right)\varphi + a= 0.
\label{f10}
\end{equation}
This confirms that the $\gamma$-dependence in the original equation was spurious and can be
trivially reconstructed from a calculation at a single value of $\gamma$.
Under these transformations, the grand canonical potential scales as follows,
\begin{eqnarray}
{\Psi_{\rm eff}} &=& \frac{\gamma}{2\pi a}\int {\rm d}\xi \left( -2a {\rm Re}\,\varphi+
(\nu^2- \sigma^2)|\varphi|^2 \right.
\nonumber \\
& & \left.  +2\nu \,{\rm Im}\,\varphi (\dot{\varphi})^* + |\varphi|^4+ |\dot{\varphi}|^2\right)
\label{f11}
\end{eqnarray}
The value of $\gamma$ only enters as an overall factor.  As a consequence, when drawing
the phase boundary in the ($\nu,\sigma$) diagram rather than the ($\mu,T$) diagram, one expects a universal curve.
Needless to say, this only gives the leading asymptotic behavior of how the tricritical
point is approached, since we truncated the Taylor expansion of the coefficients of the GL effective
action around this point after the leading terms.

We now turn to the full computation of the universal phase boundary which replaces Fig.~5c from the
variational approach. The calculation
is done differently for the 2nd and 1st order phase transition. The 2nd order phase transition can be
treated perturbatively. In this case, we set
\begin{equation}
\Phi(x)=M+\varphi(x)
\label{f11a}
\end{equation}
and keep only terms up to 2nd order in $\varphi(x)$ when computing the grand canonical potential.
Denoting the discrete Fourier components of $\varphi(x)$ by $\varphi_{\ell}$, the grand canonical potential
averaged over one period will assume the form
\begin{equation}
\frac{1}{L} \int {\rm d}x \Psi_{\rm eff} = C+ \sum_{\ell,\ell'} F_{\ell \ell'} \varphi_{\ell} \varphi_{\ell'}.
\label{f11b}
\end{equation}
Vanishing of ${\rm det}\, F$ signals the perturbative instability. $F$ being block diagonal 
(only the $F_{\ell, -\ell}$ are non-zero due to spatial averaging), it is sufficient to consider a single Fourier mode 
in order to detect the instability. Since the period is also a variational parameter, we may furthermore 
assume $|\ell| = 1$ without
loss of generality. As a result, 
the perturbative calculation is similar to the toy model of the previous section, except that one has to use 
the more general ansatz where the field $\Phi$ traces out an ellipse rather than a circle in the complex
plane,
\begin{equation}
\Phi(x) = M + A \cos(Qx) + {\rm i}B \sin(Qx).
\label{f12}
\end{equation}
We expand the effective action to 2nd order in $A,B$ and study where the homogeneous system
gets unstable. The information which we get in this manner is the boundary of this perturbative
instability, the ratio of the main axes of the ellipse, $R=A/B$, and the wavenumber $Q$ in the unstable direction
at the boundary. Let us introduce once more rescaled parameters for the mass $M$ and the wavenumber $Q$,
\begin{equation}
M=m \gamma^{1/3}, \qquad Q=q \gamma^{1/3}.
\label{f13}
\end{equation}
We then find the following parametric representation of the critical curve,
\begin{eqnarray}
2m \nu^2 & = & a + 2m^3+\sqrt{a(a+4m^3)},
\nonumber   \\
2m \sigma^2 & = & -a +6m^3+\sqrt{a(a+4m^3)}.
\label{f14}
\end{eqnarray}
The parameters $q,R$ of the unstable mode are given by
\begin{eqnarray}
q^2 & = & \nu^2 +\sigma^2 -4m^2 ,
\nonumber \\
R &=& \frac{\nu\sqrt{\nu^2+\sigma^2-4m^2}}{\nu^2+m^2}.
\label{f15}
\end{eqnarray}
The curve (\ref{f14}) is shown in Fig.~6.
\begin{figure}
\begin{center}
\epsfig{file=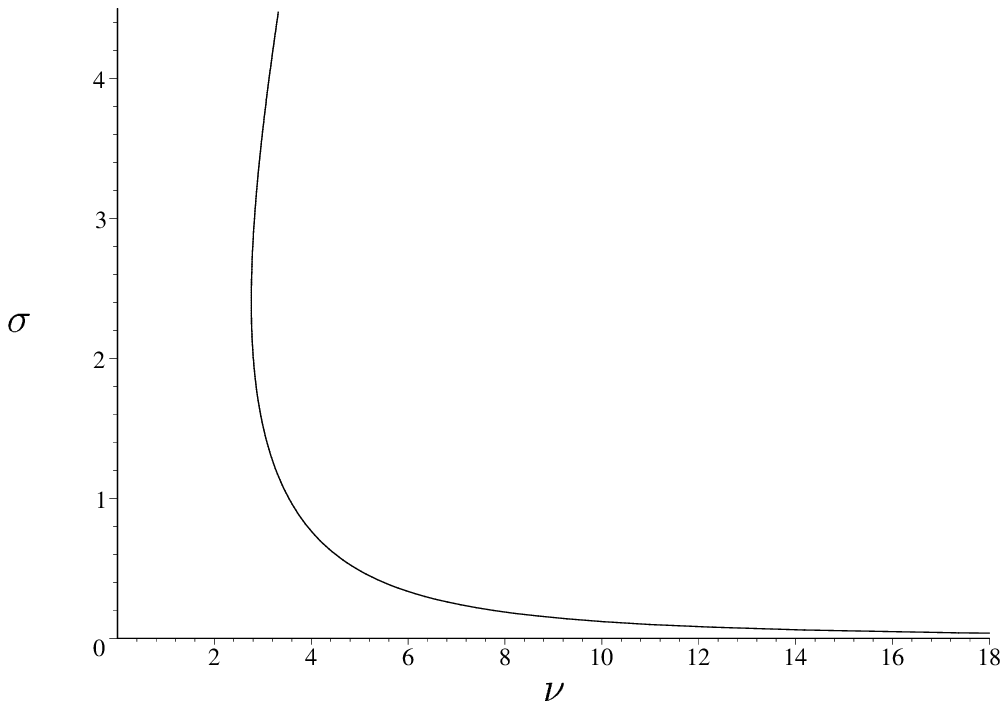,width=6.0cm,angle=270}
\end{center}
\caption{Perturbative phase boundary in the rescaled variables ($\nu,\sigma$) defined in Eq.~(\ref{f9}), computed
from the parametric representation (\ref{f14}).}
\end{figure}
This perturbative calculation does not teach us where the tricritical point is, and therefore where the 2nd order
phase boundary ends (it starts at small $\sigma$, large $\nu$, i.e., from the bottom right of the plot).
In Fig.~7 we show the ratio $R$ characterizing the eccentricity of the ellipse.
\begin{figure}
\begin{center}
\epsfig{file=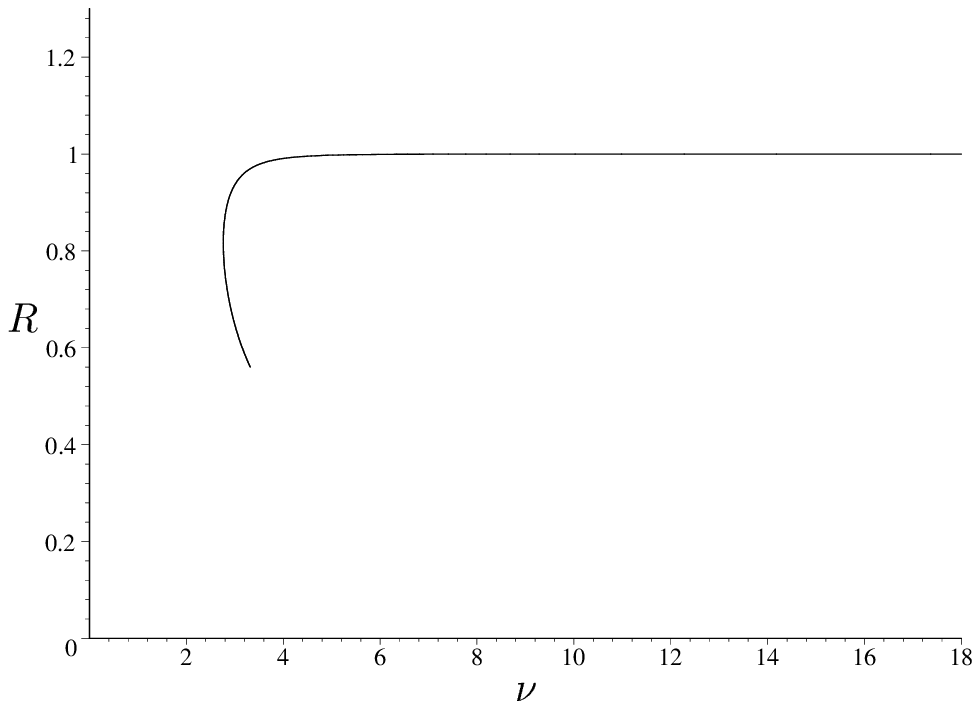,width=6.0cm,angle=270}
\end{center}
\caption{Ratio $R=A/B$ of the main axes of the ellipse traced out by $\Phi$ in the complex plane, right at the
perturbative phase boundary.}
\end{figure}
In the region of interest later on, it deviates from 1 (the value assumed in the variational calculation)
by less than $6\%$. The reduced wavenumber $q$ is equal to $\nu$ for all practical purposes so that we refrain
from showing a plot. Going back to the original variables, this perfectly agrees with the wavenumber $Q=2\mu$ of the
chiral spiral at $\gamma=0$.

We now turn to the first order transition line and the tricritical point. Here, perturbation theory is of no
help since the transition is discontinuous. We therefore have to integrate the rescaled Euler-Lagrange
equation numerically. In order to explain our method, it is useful to reinterpret the problem as one of 
classical mechanics. If we identify the spatial coordinate $\xi$ with time and $S,P$, the real and imaginary
parts of $\Phi$, with cartesian coordinates in a plane, the effective action becomes the Lagrange function
of a (planar) non-relativistic particle moving in a static electric and magnetic field, with the equation of motion
\begin{equation}
m \ddot{\vec{r}} = q\left( \vec{E}+ \dot{\vec{r}}\times \vec{B}\right).
\label{f16}
\end{equation}
The electrostatic potential $V(\vec{r}\,)$ ($\vec{E}=-\vec{\nabla} V(\vec{r}\,)$) is given by
\begin{equation}
\frac{q}{m} V(\vec{r}\,)= ax-\frac{1}{2}(\nu^2-\sigma^2)r^2 - \frac{1}{2}r^4.
\label{f17}
\end{equation}
Depending on the parameters ($\nu,\sigma$) it looks like a tilted inverted Mexican hat ($\sigma>\nu$) or a tilted sugarloaf ($\sigma<\nu$).
The inversion is due to the interchange of space and time, chiral symmetry is rotational symmetry in the plane, and the
term $ax$ arises from explicit chiral symmetry breaking.
The magnetic field is homogeneous and perpendicular to the plane of motion with magnitude proportional
to the chemical potential,
\begin{equation} 
\frac{q}{m}\vec{B}=-2\nu \vec{e}_z.
\label{f18}
\end{equation}
There are of course infinitely many solutions of Newton's equation, depending on the initial conditions. 
The homogeneous solution for $\varphi$ ($\varphi=m$) corresponds to the time independent solution where the particle
is sitting at the maximum of the potential. Since we expect $\varphi$ to be spatially periodic in the crystal phase, we search for
periodic orbits of the classical particle and minimize the action with respect to the initial conditions. We proceed as follows.
First we notice that energy 
\begin{equation}
E=\frac{m}{2} \dot{\vec{r}}\,^2 + q V(\vec{r}\,)
\label{f19}
\end{equation}
is conserved. In the original problem, this implies that a certain function of $\varphi$ and $\dot{\varphi}$ is $\xi$-independent,
\begin{equation}
\partial_{\xi} \left( |\dot{\varphi}|^2-(\nu^2-\sigma^2)|\varphi|^2- |\varphi|^4+2a {\rm Re} \varphi \right)=0.
\label{f20}
\end{equation}
For symmetry reasons, we can choose without loss of generality the initial conditions
\begin{equation}
\vec{r}\,(0)=\left(\begin{array}{c} x_0 \\ 0 \end{array}\right), \qquad \vec{v}\,(0) = \left(\begin{array}{c} 0 \\ v_0 \end{array}\right),
\label{f21}
\end{equation}
where $v_0$ can be computed from the energy,
\begin{equation}
v_0 = \sqrt{\frac{2}{m}(E-qV(x_0,0))}.
\label{f22}
\end{equation}
We then vary $x_0$ and solve the differential equations numerically until we find a periodic orbit. This requires an 
iterative procedure. The energy $E$ is treated as a variational parameter. The calculation is repeated for different energies
until the minimum of the action has been found.

We now turn to the results obtained with Maple. We find that the energy $E$ needed to minimize the action is
always very close to the maximum value of the potential. Periodic orbits are easily found at small $\sigma$ but
are increasingly difficult to find at larger $\sigma$ since one needs to fine-tune the initial coordinate $x_0$
to a very high precision.
This is due to the fact that the inverted Mexican hat shape becomes more pronounced and the particle travelling
along the rim will in general fall inside or outside before making a full turn.
Some representative orbits are shown in Fig.~8. The shape agrees very well with the circle assumed in the variational
calculation. The most conspicuous discrepancy is the behavior near the right end for large radii where the 
circle gets deformed. This happens close to the maximum of the inverted Mexican hat potential. A closer look at the
time dependence shows that the particle spends more time in this region, in contrast to the
assumed uniform motion underlying the variational calculation. The tricritical point can be rather easily identified
by the disappearance of the non-perturbative solution.
\begin{figure}
\begin{center}
\epsfig{file=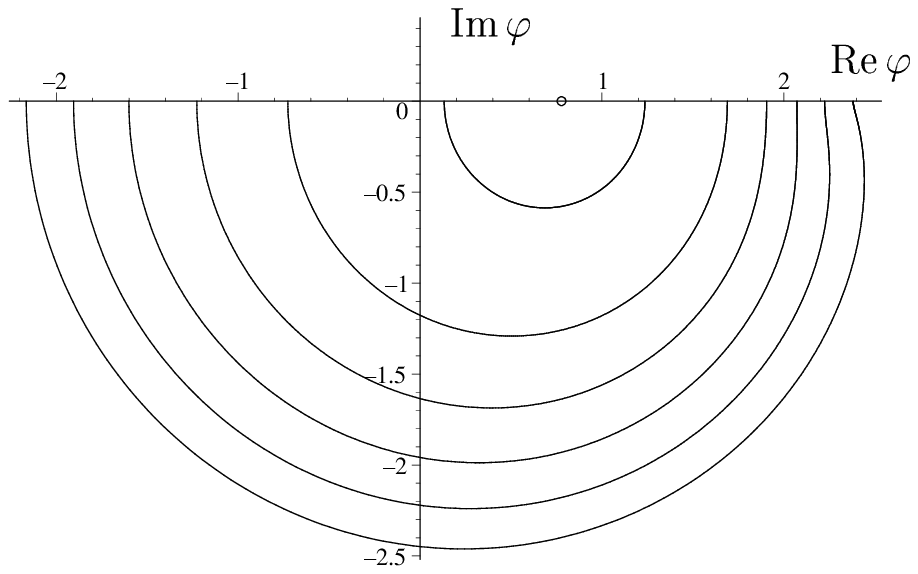,width=4.0cm,angle=270}
\end{center}
\caption{Sample periodic orbits of a classical charged particle in an electromagnetic field, equivalent to solutions of the Euler-Lagrange
equations for the rescaled GL action. The dot at $\varphi\approx 0.78$ corresponds to the time independent 
solution right at the tricritical point $(\nu,\sigma)=(2,99,1.54)$.
The curves are obtained for the parameters ($\nu,\sigma$) = (2.93,1.62), (2.58,2.15), (2.36,2.57), (2.20,2.93), (2.09,3.25), 
(1.99,3.54), from inside out.
Since the shape of the trajectories is reflection symmetric only the lower half is shown.}
\end{figure}
Fig.~9 shows the rescaled critical line. 
\begin{figure}
\begin{center}
\epsfig{file=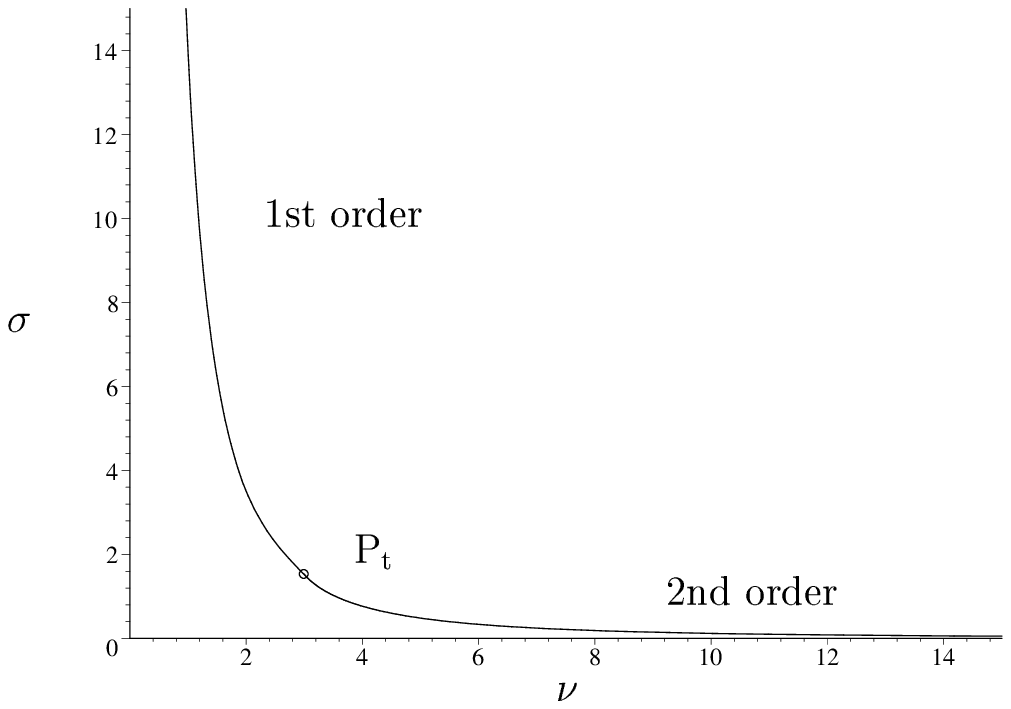,width=6.0cm,angle=270}
\end{center}
\caption{``Universal" phase boundary in rescaled variables obtained from perturbative, analytical (2nd order) and
non-perturbative, numerical (1st order) calculations.}
\end{figure}
We have combined the numerical result for the 1st order line with
the analytical calculation of the 2nd order line. The rescaled tricritical point lies at ${\rm P}_{\rm t}\approx (2.99,1.54)$,
very close to the value $(2.98,1.59)$ of the variational calculation.
As explained above, the curve shown in Fig.~9 is all what is needed to reconstruct the 3-dimensional phase
diagram in the vicinity of the tricritical point, see Fig.~10. 
\begin{figure}
\begin{center}
\epsfig{file=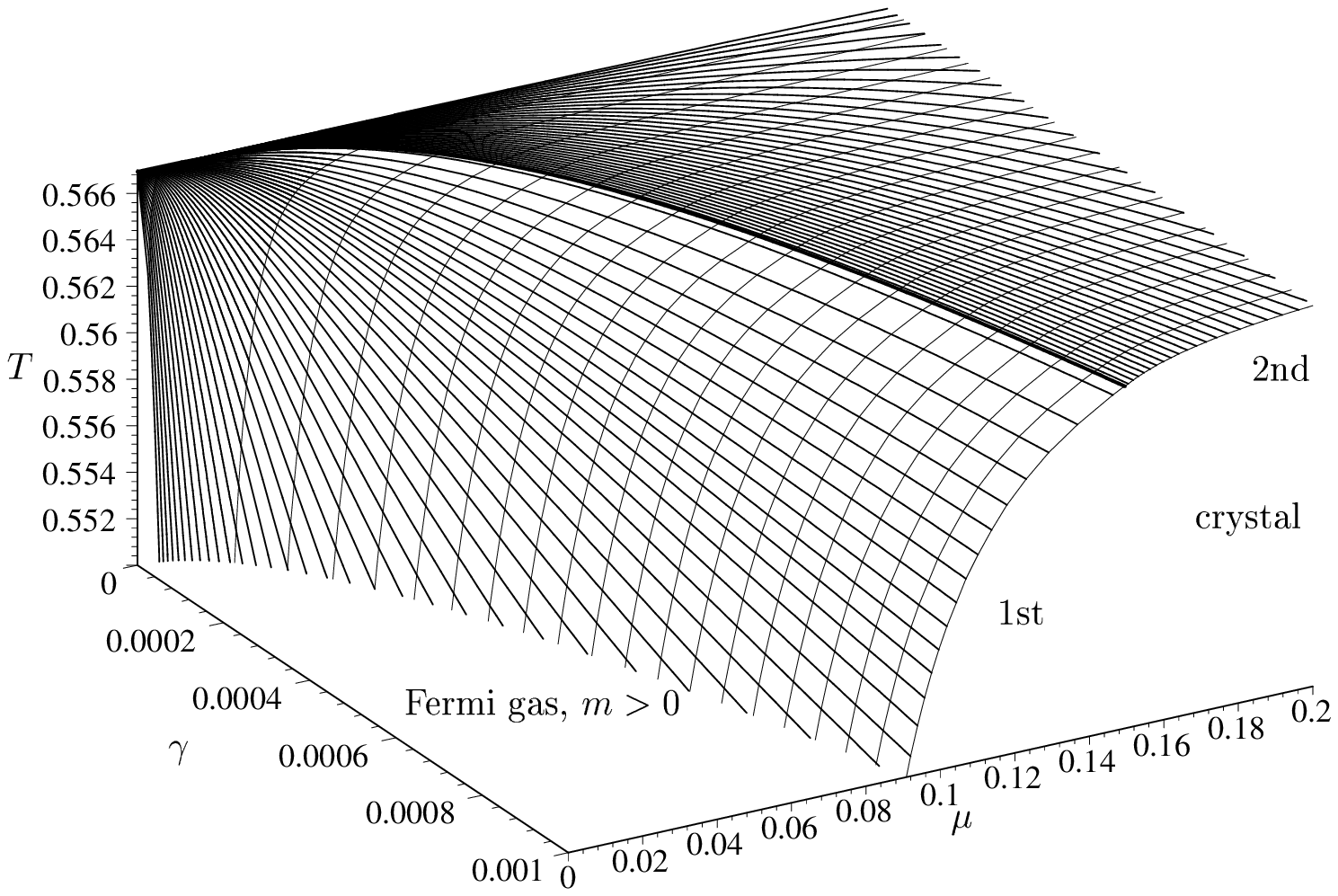,width=6.0cm,angle=270}
\end{center}
\caption{Like Fig.~4, but result from numerical minimization of the GL effective action. The whole surface
has been reconstructed from the single curve in Fig.~9 by using scaling. The curves $\gamma={\rm const.}$
would all have the same shape in a $(\mu,\tau)$ plot, but a different scale. The second set of curves drawn onto the surface 
connects points related by scaling. The narrow gap between the phase boundary and the $T$-axis
is an artefact of the limited range of $\sigma$ when determining the 1st order curve in Fig.~9.}
\end{figure}
This gives a first glimpse of the full phase diagram of the massive chiral
GN model which is still largely unknown. Although we cannot rule out further phase transitions, we believe that it contains
already the crucial information about the existing phases and the order of the phase transitions. By our derivation
it should correctly describe the asymptotics of how the tricritical point is approached as
we let $\mu\to 0,\gamma \to 0, T\to T_{\rm c}$. To illustrate this claim we mention that the tricritical line along which
the order of the phase transition changes behaves asymptotically (for $\gamma \to 0$) like
\begin{equation}
\mu  = \frac{1}{2} \nu_{\rm t} \gamma^{1/3},
\qquad
T=T_{\rm c}\left(1- \frac{1}{a} \sigma_{\rm t}^2 \gamma^{2/3} \right),
\label{f23}
\end{equation}
where now all the dynamical information is encoded in the pair of numbers $\nu_{\rm t}\approx 1.54,\sigma_{\rm t}\approx 2.99$
which had to be determined numerically. 

\section{Comparing models with discrete and continuous chiral symmetry and conclusions}
To put the results of this work into perspective, we remind the reader that the GN model with discrete chiral symmetry
and the NJL model with continuous chiral symmetry, Eq.~(\ref{a1}), were believed to have identical phase diagrams
for many years, see Fig.~2 of Ref.~\cite{9}. This was the result of the unjustified assumption that only homogeneous phases would matter.
The correct phase diagrams were first obtained for the massless models, see Fig.~1 of the present work for the NJL
model and Fig.~4 of Ref.~\cite{9} for the GN model. Subsequently the phase diagram of the massive GN model has 
been constructed (Fig.~5 of Ref.~\cite{9}). Here for fixed $\gamma$, the solitonic crystal phase is bounded by two
2nd order critical lines. They may be characterized by the following instabilities: If one lowers the temperature at 
sufficiently high chemical potential, the homogeneous condensate (fermion mass) is unstable towards a modulation
with a finite wavenumber $Q$ but vanishing amplitude. If one raises the chemical potential at sufficiently low temperature,
the system is unstable with respect to formation of a single kink-antikink baryon, i.e., a finite amplitude but vanishing
wavenumber. These two lines are joined in a cusp at a tricritical point. If one approaches the tricritical point along the first
phase boundary, $Q$ goes to zero; along the second boundary, the amplitude goes to zero. All transitions are continuous.

In the case of the massive NJL model, we do not yet have such a complete picture. 
As a first step we have presented here a study in the vicinity of the tricritical point, perhaps the most interesting
part of the full phase diagram. Again we find at fixed $\gamma$ two phase boundaries meeting at a tricritical
point. To the right of this point, we have identified a 2nd order critical line with a perturbative instability similar to
the one seen in the GN model.  Here, in addition to a spatial modulation, the mean field acquires an oscillating imaginary part
from the pseudoscalar condensate.
Another important difference is the fact that the wavenumber $Q$ of the unstable mode does not
vanish as we approach the tricritical point.  At the tricritical point, there is no cusp, but the phase boundary continues
smoothly. In contrast to the GN model it now turns into a discontinuous 1st order phase transition.  The system becomes unstable
with respect to formation of a chiral crystal with finite
period and amplitude, rather than creation of a single baryon. This is surprising, since
the massive NJL model is known to possess solitonic baryons. At $T=0$ one would definitely expect a 
2nd order transition at a chemical potential equal to the baryon mass, but apparently the transition becomes 1st order at finite
temperature. 

In conclusion, the study of the NJL thermodynamics is not just a mere repetition of that of the GN model, but involves interesting
new challenges and physics facets. Due to the complex order parameter and the fact that one can apparently do much less 
analytically, it is quite non-trivial. Nevertheless it would be interesting to determine the complete phase diagram.
Exactly soluble relativistic field theories with a rich phase structure are so rare that one should not simply
ignore this problem. Moreover, there may be applications in condensed matter physics like in the case of the GN model.
Thus for instance, relativistic four-fermion models with a continuous chiral symmetry have been 
discussed in the context of incommensurate charge-density waves some time ago \cite{16,17,18}.
Spiral order parameters have an even longer history, starting with the classic paper on spin density
waves in the electron gas by Overhauser \cite{19}. Last not least, let us mention that various chiral crystals have
been studied in 3+1 dimensional quantum chromodynamics, see e.g. the recent work \cite{20} and references therein.

\end{document}